\documentclass[unnumsec,webpdf,contemporary,large]{oup-authoring-template}%
\usepackage{color, colortbl}
\newcommand{\specialcell}[2][t]{\begin{tabular}[#1]{@{}l@{}}#2\end{tabular}}
\graphicspath{{Fig/}}

\theoremstyle{thmstyleone}%
\theoremstyle{thmstyletwo}%
\theoremstyle{thmstylethree}%

\begin{document}

\journaltitle{arXiv}
\DOI{Preprint}
\copyrightyear{2021}
\pubyear{2021}
\access{XOmiVAE}
\appnotes{Problem Solving Protocol}

\firstpage{1}

%\subtitle{Subject Section}

\title[XOmiVAE]{XOmiVAE: an interpretable deep learning model for cancer classification using high-dimensional omics data}

\author[1,2,$\dagger$]{Eloise Withnell}
\author[1,$\ast$,$\dagger$]{Xiaoyu Zhang}
\author[1]{Kai Sun}
\author[1,3]{Yike Guo}

\authormark{Zhang et al.}

\address[1]{\orgdiv{Data Science Institute}, \orgname{Imperial College London}, \orgaddress{\postcode{SW7 2AZ}, \state{London}, \country{UK}}}
\address[2]{\orgdiv{Department of Health Informatics}, \orgname{University College London}, \orgaddress{\postcode{WC1E 6BT}, \state{London}, \country{UK}}}
\address[3]{\orgdiv{Department of Computer Science}, \orgname{Hong Kong Baptist University}, \orgaddress{\state{Hong Kong}, \country{China}}}

\corresp[$\ast$]{Corresponding author. \href{x.zhang18@imperial.ac.uk}{x.zhang18@imperial.ac.uk}}
\corresp[$\dagger$]{The first two authors contributed equally to this paper.}

\received{Date}{0}{Year}
\revised{Date}{0}{Year}
\accepted{Date}{0}{Year}

%\editor{Associate Editor: Name}

%\abstract{
%\textbf{Motivation:} .\\
%\textbf{Results:} .\\
%\textbf{Availability:} .\\
%\textbf{Contact:} \href{name@bio.com}{name@bio.com}\\
%\textbf{Supplementary information:} Supplementary data are available at \textit{Briefings in Bioinformatics}
%online.}

\abstract{The lack of explainability is one of the most prominent disadvantages of deep learning applications in omics. This ``black box'' problem can undermine the credibility and limit the practical implementation of biomedical deep learning models. Here we present XOmiVAE, a variational autoencoder (VAE) based interpretable deep learning model for cancer classification using high-dimensional omics data. XOmiVAE is capable of revealing the contribution of each gene and latent dimension for each classification prediction, and the correlation between each gene and each latent dimension. It is also demonstrated that XOmiVAE can explain not only the supervised classification but the unsupervised clustering results from the deep learning network. To the best of our knowledge, XOmiVAE is one of the first activation level-based interpretable deep learning models explaining novel clusters generated by VAE. The explainable results generated by XOmiVAE were validated by both the performance of downstream tasks and the biomedical knowledge. In our experiments, XOmiVAE explanations of deep learning based cancer classification and clustering aligned with current domain knowledge including biological annotation and academic literature, which shows great potential for novel biomedical knowledge discovery from deep learning models.}

% \boxedtext{
% \begin{itemize}
% \item Key boxed text here.
% \item Key boxed text here.
% \item Key boxed text here.
% \end{itemize}}

\maketitle
\section{Introduction}
High-dimensional omics data (e.g., gene expression and DNA methylation) comprises up to hundreds of thousands of molecular features (e.g., gene and CpG site) for each sample. As the number of features is normally considerably larger than the number of samples for omics datasets, genome-wide omics data analysis suffers from the ``the curse of dimensionality'', which often leads to overfitting and impedes wider application. Therefore, performing feature selection and dimensionality reduction prior to the downstream analysis has become a common practice in omics data modelling and analysis \citep{Meng2016DimensionRT}. Standard dimensionality reduction methods like Principal Component Analysis (PCA) \citep{Ringnr2008WhatIP} learn a linear transformation of the high-dimensional data, which struggles with the complicated non-linear patterns that are intractable to capture from omics data. Other non-linear methods such as t-distributed Stochastic Neighbor Embedding (t-SNE) \citep{Maaten2008VisualizingDU} and Uniform Manifold Approximation and Projection (UMAP) \citep{McInnes2018UMAPUM} have become increasingly popular, but still have limitations in terms of scalability.

Deep learning has proven to be a powerful methodology for capturing non-linear patterns from high-dimensional data \citep{LeCun2015DeepL}. Variational Autoencoder (VAE) \citep{Kingma2014AutoEncodingVB} is one of the emerging deep learning methods that have shown promise in embedding omics data to lower dimensional latent space. With a classification downstream network, the VAE-based model is able to classify tumour samples and outperform other machine learning and deep learning methods \citep{Zhang2019IntegratedMA, Azarkhalili2019DeePathologyDM, Hira2021IntegratedMA, Zhang2021OmiEmbedAU}. Among them, OmiVAE \citep{Zhang2019IntegratedMA} is one of the first VAE-based multi-omics deep learning models for dimensionality reduction and tumour type classification. An accuracy of 97.49\% was achieved for the classification of 33 pan-cancer tumour types and the normal control using gene expression and DNA methylation profiles from the Genomic Data Commons (GDC) dataset \citep{Grossman2016TowardAS}. Similar to OmiVAE, DeePathology \citep{Azarkhalili2019DeePathologyDM} applied two types of deep autoencoders, Contractive Autoencoder (CAE) and Variational Autoencoder (VAE), with only the gene expression data from the GDC dataset, and reached accuracy of 95.2\% for the same tumour type classification task. \cite{Hira2021IntegratedMA} adopted the architecture of OmiVAE with Maximum Mean Discrepancy VAE (MMD‐VAE), and classified the molecular subtypes of ovarian cancer with an accuracy of 93.2‐95.5\%. \cite{Zhang2021OmiEmbedAU} synthesised previous models and developed a unified multi-task multi-omics deep learning framework named OmiEmbed, which supported dimensionality reduction, multi-omics integration, tumour type classification, phenotypic feature reconstruction and survival prediction. Despite the breakthrough of aforementioned work, a key limitation is prevalent among deep learning based omics analysis methods. Most of these models are ``black boxes'' with lack of explainability, as the contribution of each input feature and latent dimension towards the downstream prediction is obscured.

Various strategies have been proposed for interpreting deep learning models. Among them, the probing strategy, which inspects the structure and parameters learnt by a trained model, have been shown to be the most promising \citep{Azodi2020OpeningTB}. Probing strategies generally fall into one of three categories: connection weights-based, gradient-based and activation level-based approaches \citep{Montavon2018MethodsFI}. The connection weight-based approach sums the learnt weights between each input dimension and the output layer to quantify the contribution score of each feature \citep{Garson1991InterpretingNC, Olden2002IlluminatingT}. \cite{Way2018ExtractingAB} and \cite{Bica2019UnsupervisedGA} adopted this probing strategy to explain the latent space of VAE on gene expression data. However, the connection weight-based approach can be limited or even misleading when positive and negative weights offset each other, when features do not have the same scale, or when neurons with large weights are not activated \citep{Shrikumar2017LearningIF}. In the gradient-based approach, contribution scores (or saliency) are measured by calculating the gradient when the input are perturbed \citep{Simonyan2014DeepIC}. \cite{Diner2018DeepProfileDL} applied a gradient-based approach, Integrated Gradients \citep{Sundararajan2017AxiomaticAF}, to explain a VAE model for gene expression profiles. This approach overcomes limitations of the connection weights-based method. Despite this, it is inaccurate when small changes of the input do not effect the output \citep{Shrikumar2017LearningIF}. The activation level-based approach conquers these drawbacks by comparing the feature activation level of an instance of interest and a reference instance \citep{Azodi2020OpeningTB}. An activation level-based  based method named Layer-wise Relevance Propagation (LRP) has been used to explain a deep neural network for gene expression \citep{Hanczar2020BiologicalIO}. Nevertheless, LRP can produce incorrect results with model saturation \citep{Shrikumar2017LearningIF}. Deep SHAP \citep{Lundberg2017AUA}, which applies the key principles from DeepLIFT \citep{Shrikumar2017LearningIF}, has been used in a variety of biological applications \citep{Tasaki2020DeepLD, Lemsara2020PathMEPB}. However, there is a lack of research on the application of Deep SHAP to interpret the latent space of VAE models and the VAE-based cancer classification using gene expression profiles.

Here we proposed explainable OmiVAE (XOmiVAE), a VAE-based explainable deep learning omics data analysis model for low dimensional latent space extraction and cancer classification. OmiVAE took advantage of Deep SHAP \citep{Lundberg2017AUA} to provide the contribution score of each input molecular feature and omics latent dimension for the cancer classification prediction. Deep SHAP was selected as the interpretation approach of XOmiVAE due to its ability to provide more accurate explanations over other methods, which likely provides better signal-to-noise ratio in the top genes selected. With XOmiVAE, we are able to reveal the contribution of each gene towards the prediction of each tumour type using gene expression profiles. XOmiVAE can also explain unsupervised tumour type clusters produced by the VAE embedding part of the deep neural network. Additionally, we raised crucial issues to consider when interpreting deep learning models for tumour classification using the probing strategy. For instance, we demonstrate the importance of choosing reference samples that makes biological sense and the limitations of the connection weight-based approach to explain latent dimensions of VAE. The results generated by XOmiVAE were fully validated by both biomedical knowledge and the performance of downstream tasks for each tumour type. XOmiVAE explanations of deep learning based cancer classification and clustering aligned with current domain knowledge including biological annotation and literature, which shows great potential for novel biomedical knowledge discovery from deep learning models.

\section{Methods}

\begin{figure*}[t!]
    \centering
    \includegraphics[width = 1\hsize]{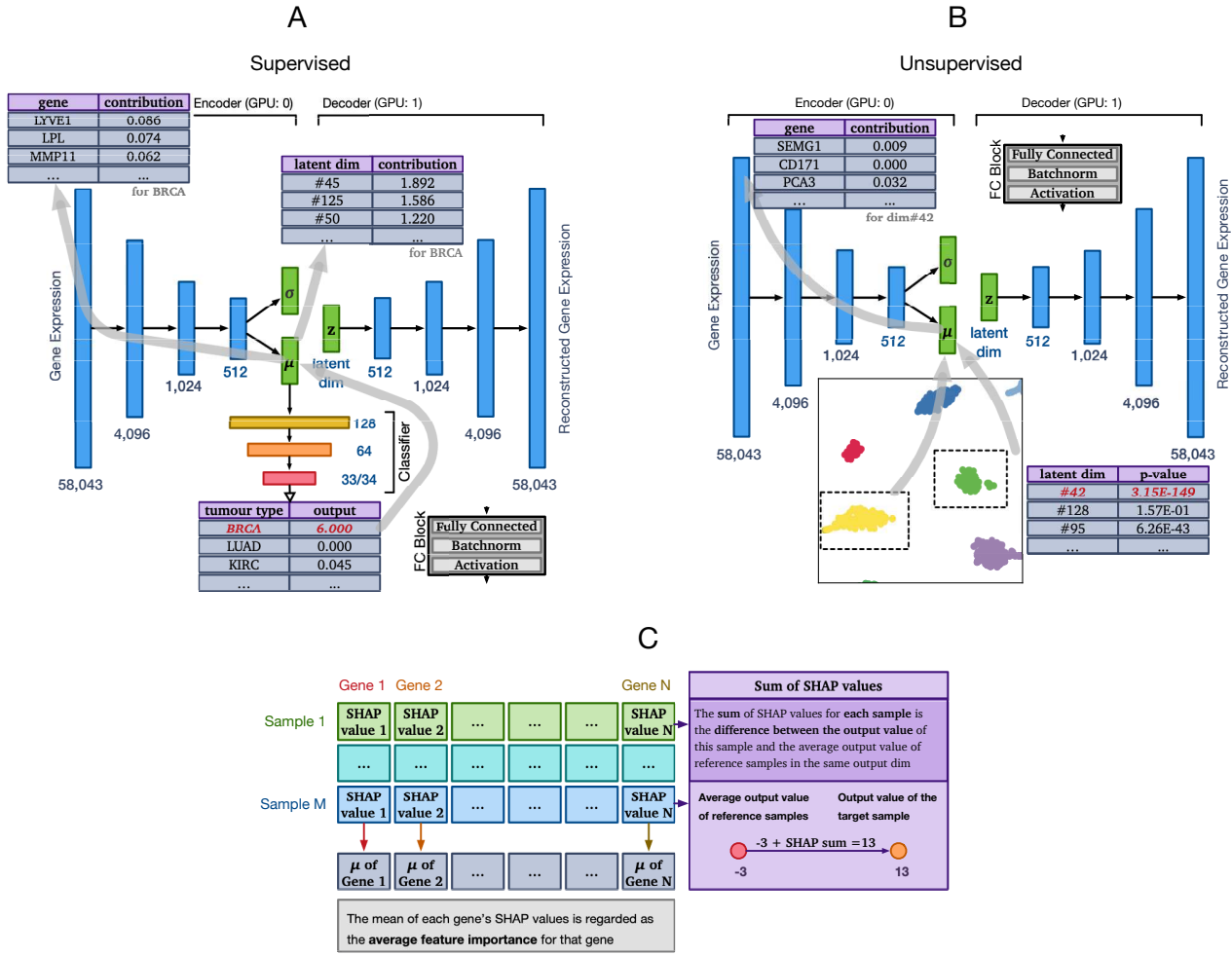}
    \caption{(A) Overall architecture of the XOmiVAE model in the supervised scenario. We can reveal the contribution score of each gene towards each cancer classification, the contribution score of each omics latent dimension learnt by VAE towards each cancer classification, and the contribution score of each gene towards each omics latent dimension. The output values and contribution scores listed in the tables are just for demonstration. (B) Overall architecture of the XOmiVAE model in the unsupervised scenario. The importance of each omics latent dimension for separating two selected clusters can be obtained using the Welch’s t-test. The contribution score of each gene can be revealed by the Deep SHAP explanation approach. The p-values and contribution scores listed in the tables are just for demonstration. (C) Illustration of how to appraise the contribution score of each gene. SHAP values were calculated for multiple samples of interest and then averaged to provide the average feature importance for each gene. To the right, we demonstrate that the SHAP values for each sample among different genes sum up to the difference between the average output value of the reference samples and the output value of the sample of interest on the same output dimension, which is another representation of the ``summation-to-delta'' property.}
    \label{fig:network_structure}
\end{figure*}

\subsection{Datasets and Pre-processing}
The Cancer Genome Atlas Program (TCGA) \citep{Weinstein2013TheCG} pan-cancer dataset, which comprise gene expression profiles of 33 various tumour types, was used in the experiment as a example to demonstrate the explainability of XOmiVAE. A total of 9,081 samples from TCGA were selected for training and testing our proposed model, of which 407 were normal tissue samples. The TCGA dataset was downloaded from UCSC Xena data portal \citep{Goldman2020VisualizingAI} (\url{https://xenabrowser.net/datapages/}, accessed on 1 May 2019). We followed the same omics data pre-processing step as OmiVAE \citep{Zhang2019IntegratedMA} and OmiEmbed \citep{Zhang2021OmiEmbedAU}. Genes targeting the Y chromosome, genes with zero expression level in all samples, and genes with missing values (N/A) in more than 10\% of the samples were removed to ensure the gene expression data was fair and clean across samples. Furthermore, the remaining N/A values that did not reach the 10\% threshold were replaced by the expression level of corresponding genes. The Fragments Per Kilobase of transcript per Million mapped reads (FPKM) values were normalised to the unit interval of 0 to 1 to the meet input requirement of the network. The phenotype data of each sample was also downloaded from UCSC Xena, which is comprised of age and gender of the subjects and primary site and disease stage of the samples. The the detailed cancer subtype information of each tumour sample was obtained from \cite{SnchezVega2018OncogenicSP}.

\subsection{Explainable OmiVAE (XOmiVAE)}
Based on vanilla OmiVAE, we proposed an interpretable deep learning model for cancer classification using high-dimensional omics data, named Explainable OmiVAE, aka XOmiVAE. The overall architecture of XOmiVAE was illustrated in Figure~\ref{fig:network_structure}. The input omics data, which was genome-wide gene expression profiles here, was first passed through a VAE embedding network to reduce the dimensionality of the input data from 58,043 to 128. The encoder of the embedding network contained two output vector, the mean vector $\boldsymbol{\mu}$ and the standard deviation vector $\boldsymbol{\sigma}$, which defined the Gaussian distribution $\mathcal{N}\left(\boldsymbol{\mu}, \boldsymbol{\sigma}\right)$ of the latent variable $\mathbf{z}$ given the input omics data $\mathbf{x}$. In order to enable backpropagation for the sampling step, the reparameterisation trick was applied according to Equation~\ref{eq:reparameterisation}:
\begin{equation}
    \mathbf{z}=\boldsymbol{\mu}+\boldsymbol{\sigma} \boldsymbol{\epsilon}
    \label{eq:reparameterisation}
\end{equation}
where $\boldsymbol{\epsilon}$ is a random variable sampled from a unit Gaussian distribution $\mathcal{N}(\mathbf{0}, \mathbf{I})$. The VAE network of XOmiVAE was optimised by maximising the variational Evidence Lower BOund (ELBO) defined in Equation \ref{eq:lower_bound}:
\begin{equation} 
    \mathrm{ELBO}=\mathbb{E}_{q_{\phi}(\mathbf{z}|\mathbf{x})}\left[-\log q_{\boldsymbol{\phi}}(\mathbf{z}|\mathbf{x})+\log p_{\boldsymbol{\theta}}(\mathbf{x}, \mathbf{z})\right] .
    \label{eq:lower_bound}
\end{equation}
$q_{\phi}(\mathbf{z|x})$ is the variational distribution introduced to approximate the true posterior distribution $p_\theta(\mathbf{z|x})$, where $\phi$ is the set of learnable parameters of the encoder and $\theta$ is the set of learnable parameters of the decoder. Equation \ref{eq:lower_bound} can further transform to Equation \ref{eq:lower_bound_kl}:
\begin{equation} 
    \mathrm{ELBO}=\mathbb{E}_{\mathbf{z} \sim q_{\phi}(\mathbf{z|x})} \log p_{\theta}(\mathbf{x|z}) - D_{\mathrm{KL}}\left(q_{\phi}(\mathbf{z|x}) \| p_\theta(\mathbf{z})\right)
    \label{eq:lower_bound_kl}
\end{equation}
where $p_\theta(\mathbf{x|z})$ is the conditional distribution, and $D_{\mathrm{KL}}$ is the Kullback--Leibler (KL) divergence between two probability distributions.

A three-layer classification neural network was attached to the bottleneck layer of the VAE deep embedding network for the tumour type classification downstream task. The latent vector $\boldsymbol{\mu}$ was fed to the classification network as the input and passed through two hidden layers with 128 neurons and 64 neurons respectively before the probability of each tumour type was obtained by the softmax activation function in the output layer. We defined the loss function of the classification network as the cross-entropy between the ground-truth tumour type $y$ and predicted tumour type $y^{\prime}$, as shown in Equation \ref{eq:loss_class}:
\begin{equation}
    \mathcal{L}_{class} = CE(y,y^{\prime})
    \label{eq:loss_class}
\end{equation}
Thus, the overall loss function of the whole model was a weighted combination of the VAE loss $\mathcal{L}_{VAE}$ and the classification loss $\mathcal{L}_{class}$, which was defined in Equation \ref{eq:loss_all}:
\begin{equation} 
    \mathcal{L}_{total}=\alpha \mathcal{L}_{VAE} + \beta \mathcal{L}_{class}
    \label{eq:loss_all}
\end{equation}
$\alpha$ and $\beta$ weighted the two losses during training. The hyper-parameters used to train this model was listed in Table \ref{tab:hyper}.

\begin{table}[t!]
    \caption{Hyper-parameters used in the model.\label{tab:hyper}}
    \tabcolsep=0pt
    \begin{tabular*}{\columnwidth}{@{\extracolsep{\fill}}ll@{\extracolsep{\fill}}}
        \toprule
        Hyper-parameter & Value \\ \midrule
        Latent dimension & 128 \\
        Learning rate & 1e-3 \\
        Batch size & 32 \\
        Epoch number - unsupervised & 50 \\
        Epoch number - supervised & 100 \\ 
        \botrule
    \end{tabular*}
\end{table}

XOmiVAE has the ability to explain both the supervised tumour type classification results, which was illustrated in Figure \ref{fig:network_structure} (A), and the unsupervised omics data clustering results, which was illustrated in Figure \ref{fig:network_structure} (B). Based on the vanilla OmiVAE, we integrated the Deep SHAP explanation approach to XOmiVAE in a customised way. Deep SHAP inherited the key principle from DeepLIFT, which is the ``summation-to-delta'' property. This property means that the sum of the attributions over the input equals the difference-from-reference of the output \citep{Shrikumar2017LearningIF}, which can be formalised by Equation \ref{eq:sum_delta}:
\begin{equation}
    \sum_{i=1}^{n} C_{\Delta x_{i} \Delta o} =\Delta o
    \label{eq:sum_delta}
\end{equation} 
where $\Delta o$ is the difference between the output of the reference sample and the output of the target sample, which is $\Delta o = f(\mathbf{x})-f(\mathbf{r})$, $\mathbf{x}$ is the target gene expression profile, $\mathbf{r}$ is the reference gene expression profile, $\Delta x_{i} = x_{i}-r_{i}$, $i$ is the gene index and $n$ is the number of genes used in the experiment \citep{Lundberg2017AUA}. Another representation of the ``summation-to-delta'' property was demonstrated in Figure \ref{fig:network_structure} (C). This property enables the calculation of the Shapley values, which indicate how to allocate contribution of each prediction result among the input features. Larger Shapley values, therefore, represent more important genes for the prediction of certain tumour type.

As for the implementation, a trained network was first passed to the Deep SHAP explainer object of XOmiVAE alongside the reference values to calculate the SHAP values from. The computation graph of the model was then able to effectively guide the explainer through the network to calculate the activation of neurons according to principles used by DeepLIFT. The original Deep SHAP was also modified to ensure it could take either the latent vector or the classification output vector as the output values for the contribution analysis. As recommended by \cite{Shrikumar2017LearningIF}, we used the pre-activation output instead of the post-softmax probabilities to calculate feature contribution scores.For each prediction, $n$ SHAP values corresponding to $n$ genes or $n$ latent dimensions were calculated to determine the contribution. The absolute values of SHAP values for each feature were averaged over a group of samples with the same label to indicate the overall contribution for each feature, as shown in Figure \ref{fig:network_structure} (C). This avoids the issue of positive and negative SHAP values offsetting each other when they were averaged across samples. To reveal the contribution of each omics latent dimension in unsupervised tasks, we calculated the mean and standard deviation of the latent vector values ($\boldsymbol{\mu}$ values) for the two groups of samples and applied a Welch’s t-test to obtain the most statistically significant dimension that separates the two groups. The correlation between the each latent variable and each gene was obtained by backpropagating the latent vector through the Deep SHAP explainer object of XOmiVAE.

\subsection{Bioinformatics analysis}
To evaluate the contribution results obtained by XOmiVAE, we compared genes with high contribution scores with the differentially expressed genes (DEGs) between normal and cancer samples for each tumour type. We used an R Bioconductor package TCGAbiolinks \citep{Colaprico2016TCGAbiolinksAR} to conduct the differential gene expression analysis. The DEGs were selected according to the cut-off of $0.05$ for the False Discovery Rate (FDR) adjusted p-value and the threshold of $3$ for the absolute $log_2$ fold change. To reveal the biological implication of the top genes with high contribution scores, we used the Broad Institute's Gene Set Enrichment Analysis (GSEA) software to perform pathway enrichment analysis \citep{Subramanian2005GeneSE}. Additionally, we used the curated gene sets from online databases including the Gene Ontology (GO) \citep{Harris2004GeneOC}, Kyoto Encyclopedia of Genes and Genomes (KEGG) \citep{Kanehisa2000KEGGKE}, and the Reactome pathway database (Reactome) \citep{Fabregat2018ReactomeGD} to test subtypes pathways. g:Profiler \citep{Raudvere2019gProfilerAW} was used to obtain and visualise the top pathways. GeneCard \citep{Stelzer2016TheGS} was used to obtain the specific gene set for each TCGA tumour type.

\section{Results and discussion}
\subsection{Multi-level explanation of XOmiVAE}

\begin{figure}[t!]
    \includegraphics[width = 1\hsize]{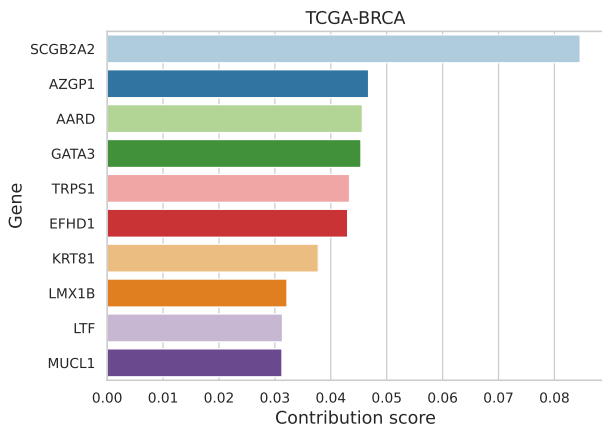}
    \centering
    \caption{The top 10 genes for the prediction of breast invasive carcinoma (BRCA). Random samples were used as the reference.}
    \label{fig:BRCA_top}
\end{figure}

\subsubsection{Most important genes for cancer classification}
We trained XOmiVAE on the TCGA pan-cancer dataset and calculated the contribution score of each gene for the prediction of each tumour type. The model achieved high accuracy for differentiating between normal and tumour tissue. For instance, the classification accuracy of BReast invasive CArcinoma (BRCA) and normal breast tissue was 99.6\% and 100\% respectively. The contribution scores followed a power-law distribution, which suggested the majority of input features (i.e., genes) were unimportant for cancer prediction (see Supplementary Figure S1-S2). As an example, we illustrated the top 10 genes with the highest scores that contributed most to the BRCA prediction in Figure \ref{fig:BRCA_top}. This demonstrated the explainability of XOmiVAE by revealing the contribution of each input feature (i.e., gene). To validate whether the top genes found by XOmiVAE made biological sense, we analysed the biological function of the most and least important genes. The top genes are known to be related to BRCA. For example, the top 1 gene for BRCA, \textit{SCGB2A2}, which codes for the protein Mammaglobin A, is highly specific of breast tissue and increasingly being used as a marker for breast cancer \citep{Lacroix2006SignificanceDA}. The second most important gene, \textit{AZGP1}, is associated with an aggressive breast cancer phenotype \citep{Parris2014AdditiveEO}. On the contrary, the 20 least important genes are either non-coding RNAs or pseudogenes with minor biological function, which are reasonable to be irrelevant when distinguishing breast tumour from normal breast tissue. A list of top genes with their contribution scores for the other 32 tumour types was also obtained by XOmiVAE and shown in Supplementary Figure S3 to S6. 

\subsubsection{Most important dimensions for cancer classification}
By passing an interim layer to the Deep SHAP explainer object of XOmiVAE, it is possible to obtain the most important neuron for a prediction in a specific layer. In the case of OmiVAE, the most intriguing interim layer to explain is the bottleneck layer, where the high dimensional gene expression data is reduced into a latent representation with lower dimensionality, 128 dimensions in our scenario. Therefore, the input of the first layer in the classification network was intercepted and explained using XOmiVAE. As an example, we show the top dimensions for different subtypes of kidney tumours: kidney chromophobe (KICH), kidney renal clear cell carcinoma (KIRC) and kidney renal papillary cell carcinoma (KIRP). The top two dimensions are different among kidney tumour subtypes and the third one was shared (Table \ref{tab:kidney_dimension}), which is therefore possible the dimension responsible for separating the kidney located tumours. Additionally, it is practicable to find the most associated genes and, therefore, the most related biological pathways to a specific dimension. We investigate the top 15 genes for the shared kidney dimension 35 as an example (see Supplementary Figure S12). These results can be obtained for every dimension and every tumour type.

\begin{table}[t!]
    \caption{The top dimensions for kidney tumour subtypes: KICH, KIRC and KIRP.\label{tab:kidney_dimension}}
    \tabcolsep=0pt
    \begin{tabular*}{\columnwidth}{@{\extracolsep{\fill}}cccc@{\extracolsep{\fill}}}
        \toprule
        & \multicolumn{3}{c}{Kidney cancer subtypes} \\ \cline{2-4}
        Dimension rank & KICH         & KIRC         & KIRP        \\ \midrule
        1st         & 45           & 20           & 42          \\ \midrule
        2nd         & 50           & 83           & 67          \\ \midrule
        \textbf{3th}         & \textbf{35}           & \textbf{35}           & \textbf{35}          \\ \midrule
        4th         & 111          & 53           & 125         \\ \midrule
        5th         & 42           & 103          & 45          \\ \botrule
    \end{tabular*}
\end{table}

\subsection{Validation by biomedical knowledge}
\subsubsection{Biomedical meaning of the top genes}
\begin{figure}[t!]
    \includegraphics[width = 1\hsize]{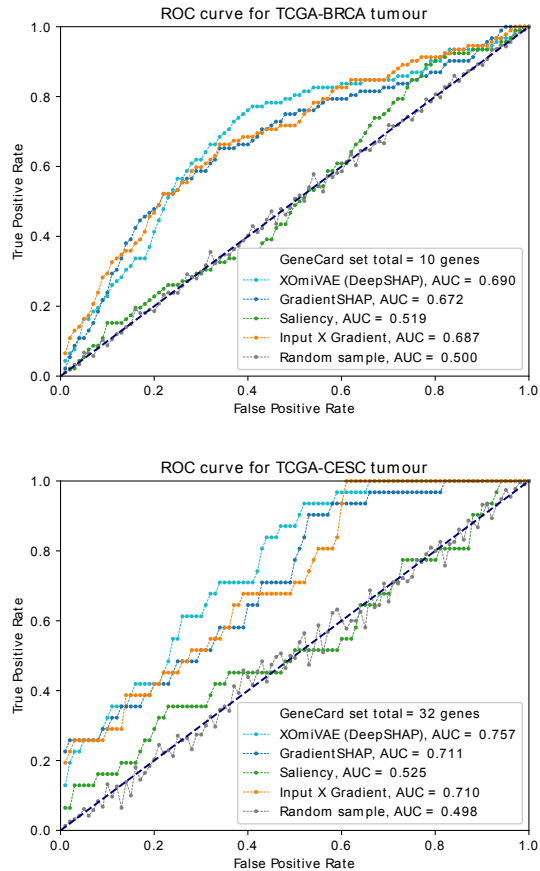}
    \centering
    \caption{AUC-ROC curves of genes as ranked by the XOmiVAE importance scores, for breast invasive carcinoma (BRCA) and cervical squamous cell carcinoma and endocervical adenocarcinoma (CESC) tumour prediction, against the GeneSet gene list for the respective tumour type. State-of-the-art methods (i.e., Saliency, Input X Gradient and GradientSHAP) and a random selection of genes are used for comparison.}
    \label{fig:ROCAUC}
\end{figure}

To validate the top genes revealed by XOmiVAE, we first compared the genes, as ranked by contribution, for each tumour type, with genes associated with the corresponding tumour type from GeneCard \citep{Stelzer2016TheGS}. GeneCard was chosen due to its comprehensive disease gene sets, which are integrated from around 150 different web sources, and therefore covered the majority of tumour types in our analysis. We selected the genes to compare at 100 different thresholds, spaced evenly from 1 (the most important gene) to 58,043 (the total number of genes). XOmiVAE were compared to different thresholds of a random sample of genes (averaged over 10 random seeds) and state-of-the-art methods, including Saliency \citep{Simonyan2014DeepIC}, Input X Gradient (an extension of Saliency) and GradientSHAP \citep{Lundberg2017AUA}. The results were plotted as a ROC curve for the True Positive Rates (TPR) and False Positive Rates (FPR). The TPR were calculated by, 
\begin{equation}
\frac{\#\;of\;top\;genes\;which\;are\;GeneCard\;disease\;genes}{\#\;of\;GeneCard\;disease\;Genes}.
\end{equation}
And the FPR were calculated by,
\begin{equation}
\frac{\#\;of\;top\;genes\;which\;are\;not\;GeneCard\;disease\;genes}{\#\;of\;genes\;not\;associated\;with\;the\;GeneCards\;disease}.
\end{equation}
21 tumour types had gene sets found in GeneCard and were therefore chosen for analysis. The ROC curves and AUC metrics are shown in Supplementary Figures S7 to S9. Two example ROC curves are illustrated in Figure \ref{fig:ROCAUC}. All 33 tumour types had an AUC metric considerably higher than the random samples which suggests that the most important genes returned by XOmiVAE are biologically relevant. The average AUC metric among all 33 tumour types of XOmiVAE and three state-of-the-art methods was listed in Table \ref{tab:auc}. XOmiVAE outperformed all of the three state-of-the-art methods.

\begin{table}[t!]
    \caption{The average AUC score across all 33 tumour types for the ranked gene importance scores compared to the GeneCard genes.\label{tab:auc}}
    \tabcolsep=0pt
    \begin{tabular*}{\columnwidth}{@{\extracolsep{\fill}}lll@{\extracolsep{\fill}}}
        \toprule
        Methods & Average AUC & Standard deviation \\ \midrule
        Saliency & 0.5331 & 0.0839 \\ 
        GradientSHAP & 0.7682 & 0.0578 \\
        Input X Gradient & 0.7762 & 0.0767 \\
        XOmiVAE & \textbf{0.7950} & 0.0673 \\
        \botrule
    \end{tabular*}
\end{table}

To further explore and understand the top genes revealed by XOmiVAE, they were evaluated using gene set enrichment analysis. We used g:Profiler \citep{Raudvere2019gProfilerAW}, a web server for functional enrichment analysis, to identify the most significant GO terms enriched in the top genes for each tumour type. Supplementary Table S1 lists the GO terms that are significantly overrepresented in top BRCA genes. The most significant GO terms related to the extracellular matrix organisation, an area of focus within breast cancer research \citep{Walker2018RoleOE}. A break down of the pathways found from the other sources used in g:Profiler was shown in Supplementary Figure S10.

The top 100 most important genes for BRCA over normal breast tissue were compared with the differentially expressed genes (DEGs) between the target tumour and normal tissue. This helps ascertain the similarity between top genes found by XOmiVAE and DEGs obtained by the traditional statistical method. We find that there is an overlap of 48 out of the 100 top contribution genes when comparing BRCA versus normal breast tissue as an example (Figure \ref{fig:DEGS}). The top DEGs were chosen according to the threshold of $FDR<0.05$ and $|LogFC| >=3$ (see Supplementary Table S2 for details). The top genes obtained by XOmiVAE do not solely include DEGs, likely because the model has to ensure that the genes chosen for classification are different between cancers. Therefore, the DEGs that are common between cancers are not chosen as important features.

\begin{figure}[t!]
    \includegraphics[width = 0.9\hsize]{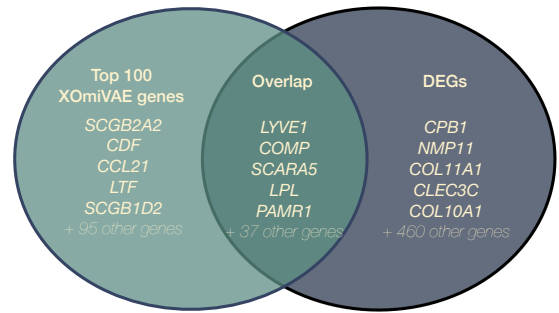}
    \centering
    \caption{A Venn diagram representing the overlap between the DEGs and top contribution genes, highlighting a total of 42 DEGs found in the top 100 contribution genes.}
    \label{fig:DEGS}
\end{figure}

\subsubsection{Biomedical meaning of important dimensions}

To further understand the most important dimensions involved in tumour prediction, we analysed the biological meaning of the key genes used by the dimensions. As an example, we analyse the highest shared dimension (i.e. dimension 35) in the kidney cancers KIRC, KIRP and KICH (Supplementary Figure S12). \textit{APQ2} is the most important gene for that dimension for all three cancer subtypes, which is located in the apical cell membranes of the collecting duct principal cells in kidneys. Additionally, all of the other high ranking genes such as \textit{UMOD}, \textit{SCNN1G} and \textit{SCNN1B} are all well known genes associated with kidney functions \citep{Carney2016GeneticsER, Hanukoglu2016EpithelialSC}. As another example, we also explain dimension 42 and 73, the 1$^{st}$ and 2$^{nd}$ most important dimension for lung adenocarcinoma (LUAD) prediction respectively, as shown in Table \ref{tab:dimcomp}. The top genes were calculated using random training samples as the reference value, to show the most important genes for LUAD versus all the other sample types. We demonstrated that dimension 42 relies heavily on the immune response pathways, whilst dimension 73 relates to the developmental process, albeit with one highly significant immune response pathway. The top gene for dimension 73 is pulmonary-associated surfactant protein C (\textit{SPC}), a surfactant protein essential for lung function, and the top gene for dimension 42 is progestagen associated endometrial protein (\textit{PAEP}), an immune system modulator, both of which have been implicated in LUAD \citep{Schneider2015GlycodelinAN, Yamamoto2005SurfactantPG}.

We found that the most important input features for the latent dimensions varied according to the tumour type used for the analysis (Table \ref{tab:dimcomp}). This demonstrates a possible limitation of previous methods explaining gene expression classification networks using solely a connection weight approach, for example by \cite{Way2018ExtractingAB} and \cite{Bica2019UnsupervisedGA}, which show no specificity for different input samples and different prediction targets. Table \ref{tab:dimcomp} shows that for BRCA, dimension 42 uses the genes related to blood vessels, and dimension 73 relies on the embryonic genes. However, this contrasts with the most important pathways that these dimensions used for LUAD classification. XOmiVAE is able to capture this as it detects the activation of neurons using Deep SHAP, as opposed to solely the weights involved.

\begin{table*}[t!]
    \caption{The biological pathways enriched for dimension 42 and 73 when classifying BRCA and LUAD.\label{tab:dimcomp}}
    \tabcolsep=0pt
    \centering
    \begin{tabular*}{\textwidth}{@{\extracolsep{\fill}}clll@{\extracolsep{\fill}}}
        \toprule
        \multicolumn{1}{l}{Dimension ID} & Tumour type & GO biological process & FDR adjusted p-value \\ \midrule
        \multirow{10}{*}{42} & \multirow{5}{*}{LUAD} & Humoral immune response & \begin{tabular}[c]{@{}l@{}}$1.8\times 10^{-8}$\\ \end{tabular} \\ \cline{3-4} 
        & & Response to bacterium & $2.0\times 10^{-8}$ \\ \cline{3-4} 
        & & Response to stimulus & $2.0\times 10^{-7}$ \\ \cline{3-4} 
        & & Immune system process & $2.5\times 10^{-6}$ \\ \cline{3-4} 
        & & Response to other organism & $3.7\times 10^{-6}$ \\ \cline{2-4} 
        & \multirow{5}{*}{BRCA} & Circulatory system process & $4.7\times 10^{-7}$ \\ \cline{3-4} 
        & & Blood circulation &$1.3\times 10^{-6}$ \\ \cline{3-4} 
        & & Developmental process & $4.6\times 10^{-5}$ \\ \cline{3-4} 
        & & Regulation of blood pressure & $2.0\times 10^{-5}$ \\ \cline{3-4} 
        & & Humoral immune response & $2.5\times 10^{-5}$\\ \midrule
        \multirow{10}{*}{73}& \multirow{5}{*}{LUAD} & Response to external stimulus & $2.4\times 10^{-5}$ \\ \cline{3-4} 
        & & Response to bacterium & $4.5\times 10^{-5}$ \\ \cline{3-4} 
        & & Anatomical structure morphogenesis & $5.4\times 10^{-5}$ \\ \cline{3-4} 
        & & Tube development & $7.4\times 10^{-5}$ \\ \cline{3-4} 
        & & Response to biotic stimulus & $1.8\times 10^{-4}$ \\ \cline{2-4} 
        & \multirow{5}{*}{BRCA} & Anterior/posterior pattern specification & $1.2\times 10^{-7}$ \\ \cline{3-4} 
        & & Embryonic morphogenesis & $1.5\times 10^{-6}$ \\ \cline{3-4} 
        & & Embryo development & $2.0\times 10^{-6}$ \\ \cline{3-4} 
        & & Embryonic skeletal system morphogenesis & $2.3\times 10^{-6}$ \\ \cline{3-4} 
        & & Anatomical structure development & $2.3\times 10^{-6}$ \\ \botrule
    \end{tabular*}
\end{table*}

\begin{figure}[t!]  
    \includegraphics[width = 0.6\hsize]{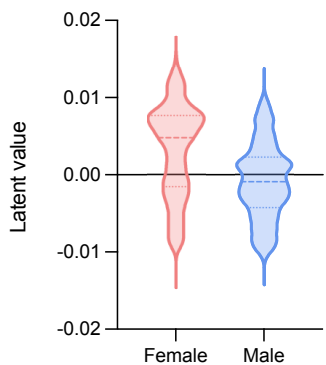}
    \centering
    \caption{Violin plot of the latent dimension 78 for female and male samples.}
    \label{fig:gender}
\end{figure}

To further understand the latent space of the classification network, we tested whether there was a dimension that separated between female and male tissue samples. We observed a large statistical difference ($p$ value = $3.6 \times 10^{-249}$) between genders on dimension 78 in the classification model (Figure \ref{fig:gender}). To understand how dimension 78 captured gender, Deep SHAP was used to explain the genes involved. We found that \textit{XIST}, a gene on chromosome X, was within the top 5 genes of the dimension 78 (Table \ref{tab:gendergenes}). \textit{XIST} is one of the key genes involved in the transcriptional silencing of one of the X chromosomes \citep{Zuccotti1995MethylationOT}. 

\begin{table}[t!]
    \caption{The top five genes for dimension 78 when separating female and male samples in the classification model of OmiVAE.\label{tab:gendergenes}}
    \tabcolsep=0pt
    \centering
    \begin{tabular*}{\columnwidth}{@{\extracolsep{\fill}}lll@{\extracolsep{\fill}}}
        \toprule
        {Gene} & {Contribution score} & {Chromosome}\\ 
        \midrule
        {CLDN3}  & \cellcolor{white!30}0.00031 & \cellcolor{white!30}chr7  \\
        {SLPI}  & \cellcolor{white!30}0.00031 & \cellcolor{white!30}chr20 \\
        {WFDC2} & \cellcolor{white!30}0.00031 & \cellcolor{white!30}chr20 \\
        \cellcolor{green!30}{XIST}  & \cellcolor{green!30}0.00030 & \cellcolor{green!30}chrX \\
        {MMP1}  & \cellcolor{white!30}0.00029 & \cellcolor{white!30}chr11 \\
        \botrule
    \end{tabular*}
\end{table}

\subsection{Validation by the performance of downstream tasks}
\subsubsection{Influence of important genes for model performance}
To further evaluate the results, we compared the classification performance of models using the top 20 XOmiVAE genes or 20 random genes for each target tumour type of interest. Four metrics including the F1-score (F1), Positive Predictive Value (PPV), True Positive Rate (TPR) and Area Under the Curve (AUC) were applied, and the performance of models using 20 random genes was averaged over 10 random seeds. A highly significant performance difference can be observed in Table \ref{tab:eval_metrics}, which indicates the contribution of the top genes obtained by XOmiVAE to the cancer classification tasks. Additionally, we calculated the average metrics for all the other tumour types except the target one and found that whilst there was also an increase in metrics from the randomly selected genes, it was not as significant as the increase for the target tumour type. This suggests that the top genes revealed by XOmiVAE are specific for certain target tumour type.

\begin{table*}[t!]
    \caption{The evaluation metrics of cancer classification using only the top 20 genes obtained by XOmiVAE (columns 1 and 3) or 20 random genes chosen from the overall gene set of 58,043 features (columns 2 and 4). The metrics for each individual tumour type of interest are shown in columns 1 and 2, and the metrics for all of the other tumour types (except the target one) are shown in columns 3 and 4. The results were averaged among all 33 target tumour types and 10 random seeds.\label{tab:eval_metrics}}
    \tabcolsep=0pt
    \centering
    \begin{tabular*}{\textwidth}{@{\extracolsep{\fill}}lcccc@{\extracolsep{\fill}}}
        \toprule
        & \multicolumn{4}{c}{Average metric across all 33 tumour types} \\ 
        \cline{2-5} 
        & \specialcell{Target tumour trained by \\ top 20 XOmiVAE genes} & \specialcell{Target tumour trained \\ by 20 random genes} & \specialcell{All other tumours trained by \\ top 20 XOmiVAE genes of \\ the target tumour} & \specialcell{All other tumours trained \\ by 20 random genes of \\ the target tumour} \\ 
        \midrule
        F1 & \textbf{${0.90}\pm{0.11}$} & \textbf{${0.46}\pm{0.21}$} & ${0.66}\pm{0.11}$ & ${0.48}\pm{0.01}$ \\
        PPV & \textbf{${0.91}\pm{0.11}$} & \textbf{${0.48}\pm{0.20}$} & ${0.69}\pm{0.08}$ & ${0.50}\pm{0.01}$ \\
        TPR & \textbf{${0.91}\pm{0.10}$} & ${0.66}\pm{0.11}$ & ${0.48}\pm{0.01}$ & ${0.48}\pm{0.01}$ \\
        AUC & \textbf{${0.94}\pm{0.07}$} & \textbf{${0.67}\pm{0.11}$} & ${0.83}\pm{0.06}$ & ${0.68}\pm{0.00}$ \\ 
        \botrule
    \end{tabular*}
\end{table*}

To approximate the most important genes for the overall model, we summed the ranking of genes for each tumour type, with the most important gene having a ranking of $1^{st}$ and the least important gene ranking $58,043^{th}$, and selected 20 genes with the lowest sum rankings to retrain the model and calculate the overall accuracy (Table \ref{tab:gene_model_acc}). We then compared it with the performance of a model trained by 20 random genes and a model trained by the overall 20 least important genes with the highest ranking sums. Using the 20 most important genes we observed a significant improvement in accuracy over using a random selection of 20 genes. Additionally, we found that the 20 least important genes caused a large decrease in accuracy compared to a random selection of genes. These results suggest a possible role of using the XOmiVAE contribution scores for feature selection in model training with high-dimensional omics data.

\begin{table}[t!]
    \caption{The accuracy of XOmiVAE using the full gene set, the top 20 contribution genes for all tumours, 20 random genes and the bottom 20 contribution genes for all tumours.\label{tab:gene_model_acc}}
    \tabcolsep=0pt
    \centering
    \begin{tabular*}{\columnwidth}{@{\extracolsep{\fill}}lccr@{\extracolsep{\fill}}}
        \toprule
        Gene set & N & Overall accuracy \\ \midrule
        Full gene set & 58,043 & ${96.85\%\pm0.46\%}$ \\
        Top 20 genes for all tumours & 20 & ${87.07\%\pm0.38\%}$ \\
        20 random genes & 20 & ${56.10\%\pm0.24\%}$ \\
        Bottom 20 genes for all tumours & 20 & ${1.68\%\pm0.37\%}$ \\ 
        \botrule
    \end{tabular*}
\end{table}

\subsubsection{Influence of important dimensions for model performance}

To understand whether XOmiVAE accurately detected the most and least important dimensions in the latent space, we evaluated the effect of knocking out the most important dimensions, Table \ref{tab:ablat}. We set the output of the target dimension to -1 when the output value was positive, and set the output of the target dimension to 1 when the output value was negative, based off a similar ablation approach by \cite{Morcos2018OnTI}. This ensures that the output is perturbed from the original value. Individually, the most important dimensions did not have a large effect when ablated, which is likely due to model saturation, a feature of neural networks that Deep SHAP addresses whereas other interpretability techniques fail to capture \citep{Shrikumar2017LearningIF}. When the top dimensions combined were ablated, the classification accuracy fell to 0. This is in contrast to the least important dimensions, which did not have any effect on the network when knocked out, individually or combined. This provides evidence to support the most and least important dimensions obtained by XOmiVAE.

\begin{table}[t!]
    \caption{The accuracy difference for each tumour type when the most important and least important dimensions were individually or together removed from the network. Values represent the mean and standard deviation of the accuracy difference among 33 tumour types.\label{tab:ablat}}
    \tabcolsep=0pt
    \centering
    \begin{tabular*}{\columnwidth}{@{\extracolsep{\fill}}ll@{\extracolsep{\fill}}}
        \toprule
        Ablated dimension & Accuracy change\\
        \midrule
        1st & $-11.9\%\pm19.9\%$ \\
        2nd & $-11.9\%\pm20.2\%$ \\
        3rd & $-2.9\%\pm6.5\%$ \\
        \textbf{Top three combined} & \cellcolor{green!30}{$-95.9\%\pm2.0\%$} \\
        \midrule
        126th & $0.0\%\pm0.3\%$ \\
        127th & $0.0\%\pm0.0\%$ \\
        128th & $0.0\%\pm0.3\%$ \\
        \textbf{Bottom three combined} & \cellcolor{green!30}{$0.0\%\pm0.3\%$} \\ 
        \botrule
    \end{tabular*}
\end{table}

\subsection{Different results depending on reference chosen}
Deep SHAP, similar to other activation level-based approaches, used reference samples as background to appraise the feature importance of each gene or latent dimension. The selection of reference samples is crucial for the explanation, since importance scores are calculated by comparing the activation level of neurons when a reference sample is fed to the network or when a target sample is fed to the network. One of the recommended choices for this reference sample is a random sample from the training set. However, we can also choose samples with certain phenotype as the reference to compare with for certain prediction rather than using a random selection of the training data, which can be more informative in some cases. For example, when explaining the important genes to differentiate gender we use samples from the opposite gender as the reference. 

\begin{figure}[t!]
    \centering
    \includegraphics[width = 0.8\hsize]{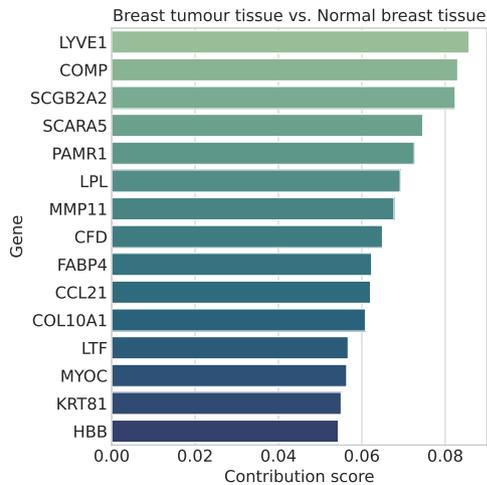}
    \caption{The top 15 genes obtained by XOmiVAE for the classification of BRCA using normal breast tissue samples as the reference.}
    \label{fig:BRCA_normal}
\end{figure}

To further understand the effect of the reference, we compared the important genes involved in BRCA classification using both a random selection from the training set and the normal breast tissue samples, Figure \ref{fig:BRCA_normal}. 25 of the top 50 XOmiVAE genes were shared between the two reference selection methods. To gain a clearer understanding of the different biological pathways enriched from the top genes when using the two different reference samples sets, we compared the g:Profiler pathway enrichment results (Supplementary Figure S10 and S11). There is a decreased enrichment of extracellular pathways when using a set of random training data to explain the BRCA predictions. As alluded earlier, extracellular pathways have been shown to be involved in BRCA progression from normal tissue \citep{Walker2018RoleOE}. It is possible that when using normal breast tissue as reference samples, the specific genes that lead to breast cancer are more pronounced, as opposed to also relying on breast tissue genes as would be the case when differentiating BRCA from all the other predictions. Therefore, it is shown that XOmiVAE is able to gain a more focused understanding of the most important genes for a tumour type by selecting the appropriate reference samples.

\subsection{Explaining unsupervised clustering results}
\begin{figure}[t!]
    \centering
    \includegraphics[width = 0.95\hsize]{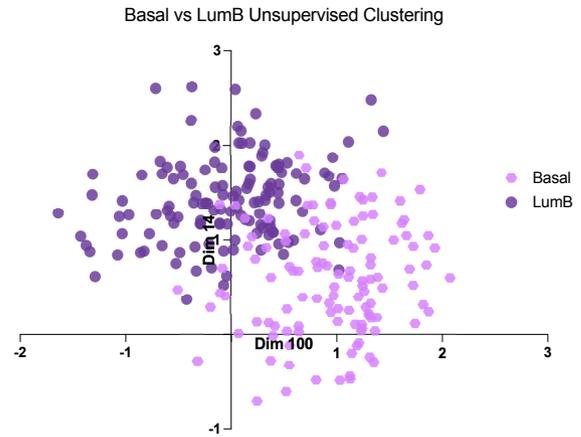}
    \caption{Top two dimensions for splitting Basal and LumB subtypes in the latent space.}
    \label{fig:subtypes}
\end{figure}

As an example of explaining the unsupervised clustering results, we used Basal-like (Basal) and Luminal B (LumB) breast tumour subtypes. Explaining the latent dimensions of VAEs would be crucial when it is important to understand the genes involved in subtype clustering of cancers that are yet to be defined, and labels that could be used for supervised learning are scarce. Figure \ref{fig:subtypes} shows the two most decisive dimensions splitting the subtypes. As the most statistically significant dimension for separating the two subtypes was dimension 100, we evaluated the enriched pathways when this dimension is used to separate Basal and LumB. Here, the $\boldsymbol{\mu}$ value for a subtype (LumB) was treated as the output and backpropagated through the network using Deep SHAP, and compared to the other subtype (Basal) as the reference. As we were interested in validating whether the model can explain the subtype specific pathways, we evaluated the top 100 genes using the Broad Institute's curated pathway database \citep{Subramanian2005GeneSE}, which includes pathways from experiments comparing the subtypes. 

In Table \ref{tab:brca_good_reference} we can see the pathways are highly specific for the subtypes. A key differentiating feature between the subtypes is that LumB is estrogen-receptor (\textit{ESR1}) positive, and Basal is \textit{ESR1} negative and in Table \ref{tab:brca_good_reference} we can see the top pathways also include the genes that differentiate between the \textit{ESR1} negative and \textit{ESR1} positive tumours. Table \ref{tab:brca_bad_reference} shows the results when the three other BRCA subtypes (LumA, Her2 and Basal) are used as the reference samples when explaining subtype LumB. The results show that a larger range of subtype pathways are present in the most important features. These results proves that it is a useful method of being able to obtain the unique genes for one subtype versus multiple other subtypes.

This is, to the best of our knowledge, the first attempt at using an activation level-based explanation approach for clustering generated by autoencoders. Typically, differential gene expression methods, such as DESeq2 \citep{Love2014ModeratedEO}, is used to explain differences in clusters, which treats each gene as independent. More recent methods improve on this, such as Global Counterfactual Explanation (GCE) \citep{Plumb2020ExplainingGO} and Gene Relevance Score (GRS) \citep{Angerer2020AutomaticIO}. However, GCE requires a linear embedding, and the embedding of GRS is constrained to ensure the gradients are easy to calculate. XOmiVAE allows for a non-linear embedding, and becomes one of the first activated-based deep learning interpretation method to explain novel clusters generated by VAEs. 

\begin{table*}[t!]
    \caption{The top pathways for differentiating LumB and Basal (BRCA subtypes), using the Broad Institute's curated pathway database.\label{tab:brca_good_reference}}
    \tabcolsep=0pt
    \centering
    \begin{tabular*}{\textwidth}{@{\extracolsep{\fill}}lcl@{\extracolsep{\fill}}}
        \toprule
        \textbf{Pathway} & \textbf{Genes in overlap} & \textbf{P-value} \\ 
        \midrule
        \specialcell{Genes up-regulated in breast cancer samples positive for ESR1 compared \\ to the ESR1 negative tumours} & 27 & 1.15e-47 \\ 
        \midrule
        \specialcell{Genes down-regulated in basal subtype of breast cancer samples} & 39 & 4.25e-43 \\ 
        \midrule
        \specialcell{Genes up-regulated in bone relapse of breast cancer} & 24 & 2.6e-42 \\ 
        \midrule
        \specialcell{Genes which best discriminated between two groups of breast cancer according \\ to the status of ESR1 and AR basal (ESR1- AR-) and luminal (ESR1+ AR+)} & 29 & 1.85e-37 \\ 
        \midrule
        \specialcell{Genes up-regulated in luminal-like breast cancer cell lines compared to the basal-like ones} & 26 & 1.82e-30 \\
        \botrule
    \end{tabular*}
\end{table*}

\begin{table*}[t!]
    \caption{The top pathways for differentiating between LumB and the other three subtypes (Basal, LumA and Her2), using the Broad Institute's curated pathway database.\label{tab:brca_bad_reference}}
    \tabcolsep=0pt
    \centering
    \begin{tabular*}{\textwidth}{@{\extracolsep{\fill}}lcl@{\extracolsep{\fill}}}
        \toprule
        \textbf{Pathway} & \textbf{Genes in overlap} & \textbf{P-value} \\ 
        \midrule
        \specialcell{Genes down-regulated in basal subtype of breast cancer samples} & 27 & 1.15e-47 \\ 
        \midrule
        \specialcell{Genes up-regulated in bone relapse of breast cancer} & 39 & 4.25e-43 \\ 
        \midrule
        \specialcell{Genes down-regulated in ductal carcinoma vs normal ductal breast cells} & 24 & 2.6e-42 \\ 
        \midrule
        \specialcell{Genes down-regulated in nasopharyngeal carcinoma (NPC) positive for LMP1, a latent gene \\ of Epstein-Barr virus (EBV)} & 29 & 1.85e-37 \\ 
        \midrule
        \specialcell{Genes up-regulated in breast cancer samples positive for ESR1 compared to the ESR1 \\ negative tumours} & 26 & 1.82e-30 \\ 
        \botrule
    \end{tabular*}
\end{table*}

\section{Conclusion}
Here we presented an explainable variational autoencoder based deep learning method for high dimensional omics data analysis, named XOmiVAE. We illustrated that it is possible to explain the supervised task of the network and obtain the most important genes and dimensions for a prediction. We also showed that it is practicable to explain the most important genes in an unsupervised network, and therefore provide a method for explaining deep learning based clustering. We evaluated the explanations of XOmiVAE and demonstrated that they make biological sense. Additionally, we offered important steps to consider when interpreting deep learning models for tumour classification. For example, we highlighted the importance of choosing reference samples that makes biological sense when explaining the model, and we disclosed the limitations of connection weight based methods to explain latent dimensions. We believe XOmiVAE is a promising methodology that could help open the “black box” and discover novel biomedical knowledge from deep learning models.

%%%%%%%%%%%%%%

\section{Key Points}
\begin{itemize}
    \item XOmiVAE is a novel interpretable deep learning model for cancer classification using high-dimensional omics data.
    \item XOmiVAE provides contribution score of each input molecular feature and omics latent dimension for each prediction.
    \item XOmiVAE is able to explain unsupervised clusters produced by VAEs without the need for labeling.
    \item XOmiVAE explanations of the downstream prediction were evaluated by biological annotation and literature, which aligned with current domain knowledge.
    \item XOmiVAE shows great potential for novel biomedical knowledge discovery from deep learning models.
\end{itemize}

\section{Availability}
The source code have been made publicly available on GitHub \url{https://github.com/zhangxiaoyu11/XOmiVAE/}. The TCGA pan-cancer dataset can be downloaded from the UCSC Xena data portal \url{https://xenabrowser.net/datapages/}.

\section{Supplementary data}
Supplementary is available at GitHub \url{https://github.com/zhangxiaoyu11/XOmiVAE/blob/main/documents/supplementary.pdf}.

\section{Acknowledgments}
This work was supported by the European Union's Horizon 2020 research and innovation programme under the Marie Sk\l{}odowska-Curie grant agreement [764281].

\section{Competing interests}
The authors declare that they have no conflict of interest.

%USE THE BELOW OPTIONS IN CASE YOU NEED AUTHOR YEAR FORMAT.
\bibliographystyle{abbrvnat}
\bibliography{reference}

%USE THE BELOW OPTIONS IN CASE YOU NEED NUMBERED FORMAT. UNCOMMENT THE ABOVE TWO LINES.
% \bibliographystyle{unsrt}
% \bibliography{reference}

%% sample for biography with author's image

%% sample for biography without author's image
\begin{biography}{}{\author{Eloise Withnell} is currently a PhD candidate at Department of Health Informatics, University College London, London, UK.}
\end{biography}

\begin{biography}{}{\author{Xiaoyu Zhang} is currently a PhD candidate at Data Science Institute, Imperial College London, London, UK.}
\end{biography}

\begin{biography}{}{\author{Kai Sun} is currently the acting operations manager of Data Science Institute, Imperial College London, London, UK.}
\end{biography}

\begin{biography}{}{\author{Yike Guo} is currently the co-director of Data Science Institute, Imperial College London, London, UK and vice-president of Hong Kong Baptist University, Hong Kong, China.}
\end{biography}

\end{document}